\documentclass{aa}  
\usepackage{amsmath}
\usepackage{graphicx}
\usepackage{txfonts}
\usepackage{natbib}
\bibpunct{(}{)}{;}{a}{}{,}
\begin{document} 

   \title{Nine new  open clusters within 500 pc from the Sun} 
   
   \author{Siegfried R\"oser
          \inst{1}
          \and
          Elena Schilbach
	  \inst{1}
          \and
	  Bertrand Goldman
	  \inst{2,3}
          }

\institute{Zentrum f\"ur Astronomie der Universit\"at
Heidelberg, M\"{o}nchhofstra\ss{}e 12-14, D--69120 Heidelberg, Germany\\
              \email{roeser@ari.uni-heidelberg.de, elena@ari.uni-heidelberg.de}
\and
{Max-Planck-Institut f\"ur Astronomie, K\"onigstuhl 17, D--69117 Heidelberg, Germany}
\email{goldman@mpia.de}
\and
{Observatoire astronomique de Strasbourg, Universit\'e de Strasbourg - CNRS UMR 7550,
 11 rue de l'Universit\'e, 67000, Strasbourg, France}
}

   \date{Received 21 June 2016; accepted xxx 2016}


  \abstract
    {}
   {One of the results of the Milky Way Star Clusters (MWSC) survey 
    by \citet{2013A&A...558A..53K} was the detection of a slight under-density of old (ca. 1 Gyr)
    clusters within the nearest kilo-parsec from the Sun. This under-density may be due to an ineffectiveness in the detection of larger structures with lower surface brightness.
     We report on our attempts to reveal such clusters.
    } 
   {We derived proper motions from a combination of Tycho-2 with URAT1, and obtained
a mean precision of about 1.4 mas/y per co-ordinate for 1.3 million stars north of -20\degr~declination.
We cut the sky into narrow proper motion slices and searched for spatial over-densities
of stars in each slice. In optical and near-infrared colour-magnitude diagrams stars from over-densities were than examined to determine if they are compatible with
isochrones of a cluster. We estimated the field star contamination using our data and the Besan\c{c}on Galactic model.}
   {We detected 9 hitherto unknown open clusters 
   in the vicinity of the Sun with ages between  70~Myr and 1~Gyr,
   and distances between 200 and 500~pc. }
   {}

   \keywords{Open clusters and associations: general --
Catalogs --
Surveys --
Proper motions --
               }

   \maketitle

\section{Introduction}
Star clusters are considered as building blocks of the Galaxy.
Their study is important for our understanding of star formation
and of the history of the Milky Way. 
The MWSC (Milky Way Star Clusters) project \citep{khea12}
aims to deliver a sample of open clusters within the larger
solar neigbourhood which covers a representative data set
of clusters of all ages. Though \citet{2013A&A...558A..53K} found that
their sample of some 2800 open clusters is essentially complete to 1.8 kpc from the Sun, for a 
small subset of the oldest open clusters ($\log t \geq 9$), they noted evidence of 
incompleteness within about 1 kpc from the Sun \citep[see Fig 4. in][]{2013A&A...558A..53K}. 

In the following years, several efforts were undertaken to fill the gap
of old clusters within 1 kpc. A classical method to reveal new open clusters is searching for 
stellar over-densities on the sky. This method has been applied by \citet{2014A&A...568A..51S}
on the basis of the 2MASS catalogue \citep{cat2MASS}. Although they found 139 new open clusters, the situation within 1 kpc 
did not change. The depth of 2MASS is responsible for a high average background noise, so this method turned out to be more effective for the 
detection of compact clusters which, in general, are the more distant ones.

For a better detection of over-densities, the background noise can be reduced by using 
proper motions to construct subsets of co-moving stars. Open clusters are gravitationally bound entities of co-eval stars,
so common motion is a necessary condition for stars forming an open cluster. The internal velocity dispersion
depends on cluster mass and reaches usually a few hundred meters per second for the vast majority of open
clusters in the Galaxy.
For example, the internal 
1-dimensional velocity dispersion found to be 0.44 km/s for an old open cluster such as the Hyades with
a tidal mass of 275  $\rm{M_\sun}$ \citep[see][]{2011A&A...531A..92R}. Even for loose ensembles such as Nearby 
Young Moving Groups the 1-dimensional velocity dispersions is about 1 km/s  
\citep[see][]{2016IAUS..314...21M}. 
At 200 pc, 1mas/y in proper motion roughly corresponds
to 1 km/s in space motion, so the internal velocity dispersion of an open cluster, and even of a moving group
should still be smaller than the typical present-day accuracy of proper motions of stars at distances larger 
than about 200 pc.

In their search for undetected clusters, 
\citet{2015A&A...581A..39S} used proper motions from a subset of 399 million stars of the \mbox{PPMXL} \citep{ppmxl} catalogue.
The accuracy of the proper motions in PPMXL
depends essentially on declination. For stars having 2MASS observations,
\citet{ppmxl} give a typical
accuracy of proper motions of 4 mas/y north of $-30^{\circ}$ declination, and 9 mas/y south of it.
Taking this into account, \citet{2015A&A...581A..39S} divided the catalogue into 
441 circular proper motion bins (radius 15~mas/yr) within $\pm$50~mas/yr in $\mu_{\alpha}\cos{\delta}$ and in $\mu_{\delta}$. 
In each proper motion bin, 
they looked for over-densities in small sky areas ('sky pixels') of 0.25$\times$0.25~deg$^2$ which corresponds 
to the typical size of clusters in the MWSC survey. As a result, 63 new open clusters were found by 
this approach, however, all were farther away from the Sun than 1 kpc.
  
Why did \citet{2015A&A...581A..39S} not find new clusters within 1~kpc?  Possibly, the missing  clusters (within 1 kpc) should have a 
larger angular extent on the sky than reflected in
the small sky pixels chosen by \citet{2015A&A...581A..39S}. Also, the large bins (a radius of 15 mas/yr) 
in proper motion space do not efficiently reduce the sky background which makes over-densities more 
difficult to be detected. For the detection of nearby clusters small (less than about 1 mas/y) proper 
motion bins and large spacial bins of about 1~deg$^2$ could be a reasonable choice. 
 
The forthcoming Gaia data releases will be ideal for this purpose. However, in this paper we try to show
that even a moderate progress in the precision of proper motions allows to reveal so far unknown open clusters
within 1 kpc from the Sun. The results presented here should be looked at as a today's  
snapshot of what we can expect from Gaia data. Note that it was not our intention to
determine the most accurate astrophysical parameters of these clusters on the basis of the
presently available astrometric and photometric data as these parameters will be revised
within a short time using the Gaia results.

In Section 2 we describe how we improved the Tycho-2 \citep{tycho2} proper motions, followed by explaining
the adopted method of cluster detection in Section 3. In Section 4 we present the newly found clusters,
and the discussion in Section 5 concludes the paper.

\section{Improving Tycho-2 proper motions}

In order to test what proper motions, more precise than those of Tycho-2,
can do for open cluster studies we have used URAT1 \citep{2015AJ....150..101Z} to improve
the Tycho-2 proper motions. URAT1 contains 228 million objects down to about $R = 18.5$~mag, 
north of about -20\degr declination. For the bulk of the Tycho-2 stars, URAT1 gives positions
at a mean epoch around 2013.5 and an accuracy level of about 20 mas per co-ordinate.
We cross-matched URAT1 with Tycho-2 (the original data set tyc2.dat from CDS), and new proper motions 
have been obtained by a least-squares adjustment as described, e.g. in \mbox{PPMXL} \citep{ppmxl}.
To avoid formally ultra-precise astrometry for a small number of stars, a 10 mas floor has been chosen for 
the precision of a URAT1 position.

We constructed a catalogue, hereafter called TYCURAT, containing 1,300,000 stars north of 
-20\degr  declination with an average formal
precision of 1.35 mas/y in proper motion, both in right ascension and in declination. The corresponding
accuracy in Tycho-2 is 2.37 mas/y, which gives a formal improvement of proper motion accuracy of 40\%.
URAT1 was astrometrically calibrated by its authors using UCAC4 \citep{2013AJ....145...44Z} which is nominally 
on the ICRS system.
No attempt has been made to investigate potential regional systematic distortions in the
proper motion system of the new catalogue TYCURAT. These possible zonal deviations are not crucial for a potential detection of open clusters and moving groups 
on areas of a few square degrees. 

TYCURAT was then cross-matched with ASCC-2.5 \citep{2001KFNT...17..409K} to obtain the Johnson $B, V$ and
2MASS photometry included in the latter \footnote{TYCURAT is available on request from
roeser@ari.uni-heidelberg.de}.
\section{Detecting nearby stellar groups.}
We note that the stellar content of TYCURAT is a subset of the ASCC-2.5 catalogue which was already used for
a systematic search of open clusters by \citet{newc109}. They
revealed 109 new Galactic open clusters by looking around each star (seed) brighter than V = 9.5 in an area 
less than 0.3 degrees,
if they could find companions with proper motions within 1-$\sigma$ of the seed's proper motion. 
In a next step, they checked if the selected stars 
fit to an isochrone in an $(B-V,V)$ colour-magnitude-diagram (CMD).
Using TYCURAT which has the same stellar content on 67 per cent of the sky, 
we could not expect to find a large number of new clusters. 
However, with improved proper motions and changing the detection 
criteria we get a chance to reveal those clusters that escaped the detection by
\citet{newc109}.
\subsection{Astrometric steps.}
In our approach, we first start with a narrow range of 1.5~mas/yr in proper motion space to down-select 
stars having the "same" proper motion and look for their concentrations
in circular bins with a radius of 0.5 degrees on the sky to reveal cores of possible co-moving objects.
The details of the procedure are described below.

a) Depending on their proper motions, we divided \mbox{TYCURAT} stars into subsets (which we call proper 
motions slices) by the following rule.
In the proper motion plane ($\mu_{\alpha}\cos{\delta}$, $\mu_{\delta}$) we select a quadratic area 
around the origin given by the condition 
$|\mu_{\alpha}\cos{\delta}| + |\mu_{\delta}| \leq 43$~mas/y
and chose grid points (centres) at each integer value $(i,k)$. This gives 3785 proper motion slices. 
A slice $(i,k)$
then contains all stars for which $(\mu_{\alpha}\cos{\delta} -i)^2+(\mu_{\delta}-k)^2 \leq 1.5^2$ holds.
The number of stars per slice is strongly varying and depends on the total proper motion. 
The maximum number of 44000 stars is at ($\mu_{\alpha}\cos{\delta}$, $\mu_{\delta} $) = \mbox{($-1,-3$)},
whereas at the borders of our square the number of stars is less than 150. About 90\% of all slices 
contain less than 6000 stars. These slices represent a seven-fold overlap
of the proper motion plane, with the effect that the same over-density on the sky may appear in neighbouring 
slices, but not necessarily with the same signal-to-noise ratio.

b) In the second step we searched for over-densities in the proper motion slices. 
For each star in a proper motion slice we counted all 
its neighbours within a radius of 0.5 degree (called {\em sky circle}) and determined the surface
density $S$ (stars/square degree). Surrounding each star's
{\em sky circle} we chose an area of 9 square degrees where we determined the background surface
density $B$. We only retain stars if $S/B$ of their {\em sky circle} is larger than
3.0 and their $S-B$ is larger than 5 per square degree. If the angle between two stars is less than 0.5~degree,
each of them is in the other's {\em sky circle}. We discard the one with the lower $S/B$. This
cleaning goes on until only 'isolated' {\em sky circles} are left, which we call
Level-0 centres.

As our system of proper motions slices represents a seven-fold overlap of the
sky, a Level-0 centre in slice $n$ could have the same or a neighbouring
Level-0 centre in slice $m$. If they are within the radius of a {\em sky circle} and the difference of their actual proper motions is less than
1.5 mas/y, only the Level-0 centre with higher $S/B$ is retained.  These
survivors are called Level-1 centres; they are disjunct and are possible candidates of new 
open clusters.

Of course, many of the Level-1 centres could be related to known open clusters. So we cross-matched them 
with the MWSC clusters \citep{2013A&A...558A..53K} and with the most recent cluster list DAML02 VERSION 3.5 of 
2016 Jan 28 \citep{daml02}.
We  discarded all the Level-1 centres within 0.7 degree of a known cluster, 
if the radius of the known cluster  is larger than 0.2 degree.
If the radius of a cluster in MWSC or in DAML is smaller than 0.2 degree, it cannot be identical to one
of the objects found here.
About two hundred Level-1 centres surviving this process 
(called Level-2 centres) were then subject to a photometric check for their existence of being clusters.
\subsection{Photometric steps.}
a) We first made a quick and coarse check in the $(J - K_S, K_S)$  CMD of the stars in a {\em sky circle} around a
Level-2 centre by using  TOPCAT\footnote{Tool for OPerations on Catalogues And Tables}\citep{2005ASPC..347...29T} .
A Level-2 centre was selected for a more careful treatment, if stars with
$(J - Ks) < 0.4$ showed a tendency for building a part of a main sequence or a turn-off region. 
The Level-2 centres were immediately rejected, if, on the CMDs, their stars showed a random distribution, mostly at the limiting magnitude of Tycho-2.
So, we could quickly check the \mbox{Level-2} centres found in the previous step.
\begin{figure*}[ht!]
\begin{minipage}[t]{0.250\textwidth}\vspace{0pt}
\includegraphics[width=\textwidth]{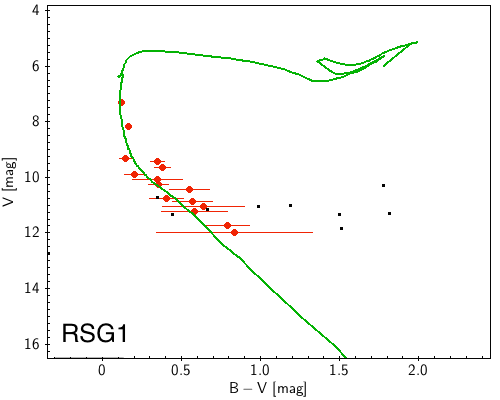}
\end{minipage}\hfill%
\begin{minipage}[t]{0.250\textwidth}\vspace{0pt}
\includegraphics[width=\textwidth]{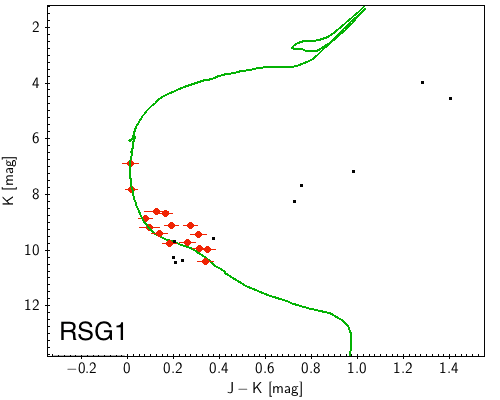}
\end{minipage}\hfill%
\begin{minipage}[t]{0.250\textwidth}\vspace{0pt}
\includegraphics[width=\textwidth]{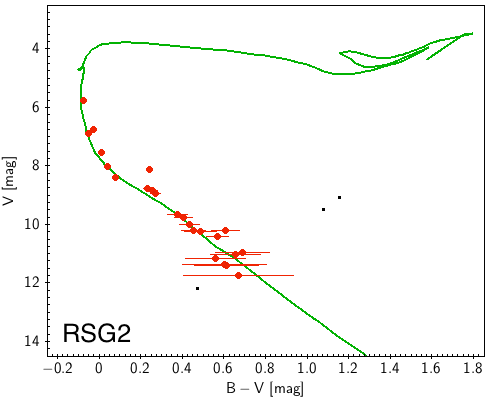}
\end{minipage}\hfill%
\begin{minipage}[t]{0.250\textwidth}\vspace{0pt}
\includegraphics[width=\textwidth]{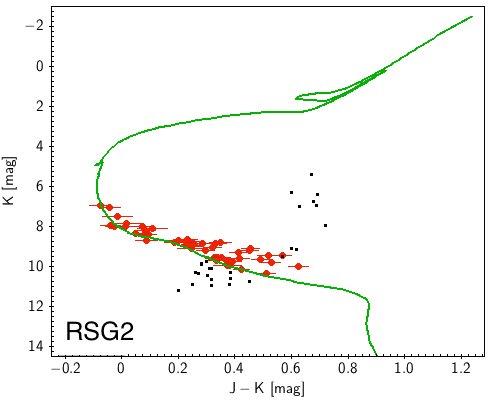}
\end{minipage}\\[5pt]%
\begin{minipage}[t]{0.250\textwidth}\vspace{0pt}
\includegraphics[width=\textwidth]{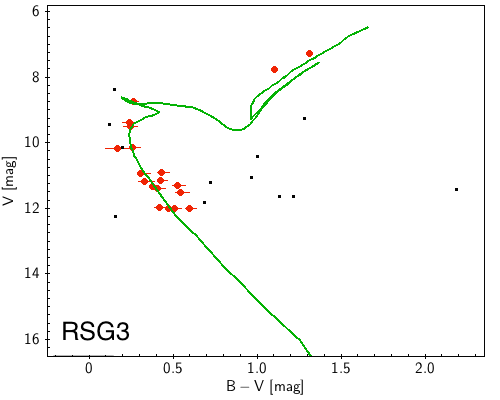}
\end{minipage}\hfill%
\begin{minipage}[t]{0.250\textwidth}\vspace{0pt}
\includegraphics[width=\textwidth]{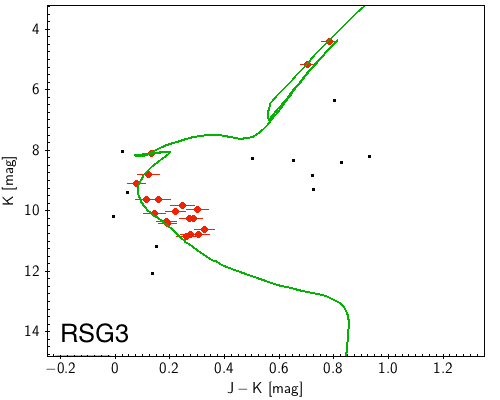}
\end{minipage}\hfill%
\begin{minipage}[t]{0.250\textwidth}\vspace{0pt}
\includegraphics[width=\textwidth]{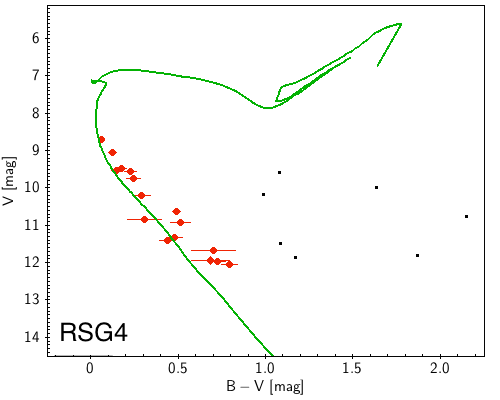}
\end{minipage}\hfill%
\begin{minipage}[t]{0.250\textwidth}\vspace{0pt}
\includegraphics[width=\textwidth]{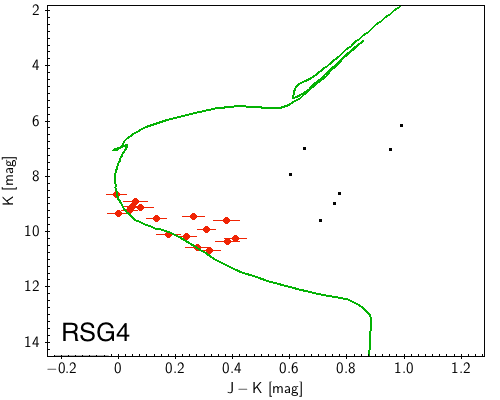}
\end{minipage}\\[5pt]%
\begin{minipage}[t]{0.250\textwidth}\vspace{0pt}
\includegraphics[width=\textwidth]{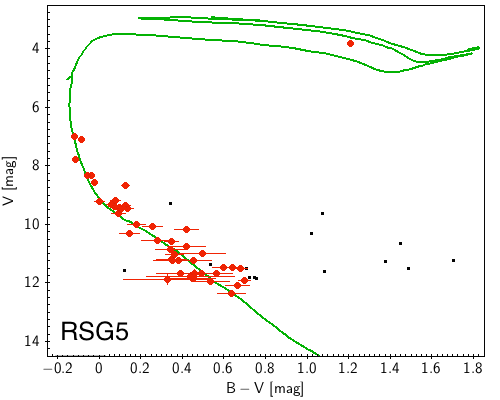}
\end{minipage}\hfill%
\begin{minipage}[t]{0.250\textwidth}\vspace{0pt}
\includegraphics[width=\textwidth]{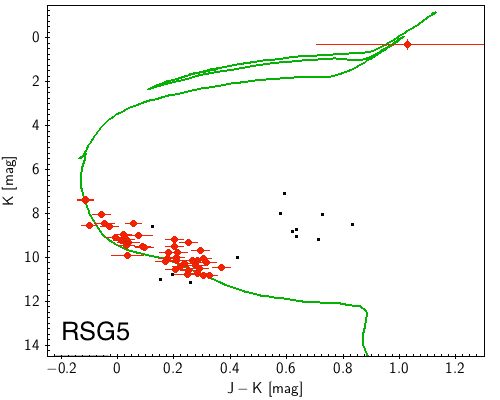}
\end{minipage}\hfill%
\begin{minipage}[t]{0.250\textwidth}\vspace{0pt}
\includegraphics[width=\textwidth]{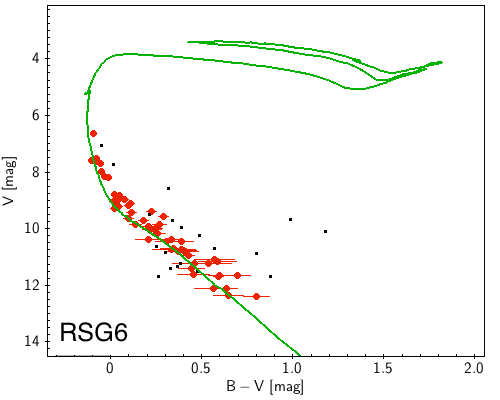}
\end{minipage}\hfill%
\begin{minipage}[t]{0.250\textwidth}\vspace{0pt}
\includegraphics[width=\textwidth]{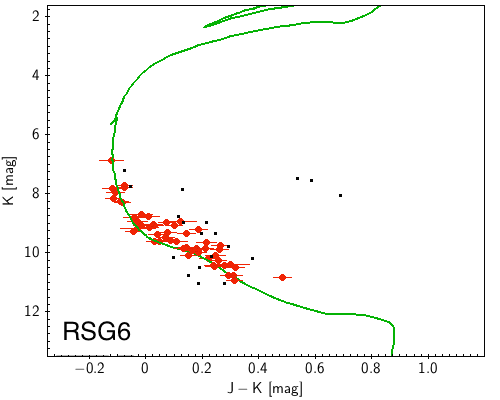}
\end{minipage}\\[5pt]%
\begin{minipage}[t]{0.250\textwidth}\vspace{0pt}
\includegraphics[width=\textwidth]{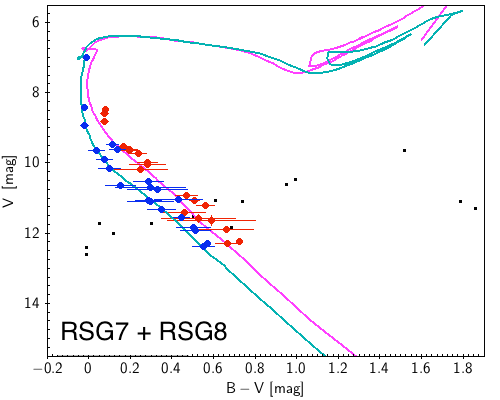}
\end{minipage}\hfill%
\begin{minipage}[t]{0.250\textwidth}\vspace{0pt}
\includegraphics[width=\textwidth]{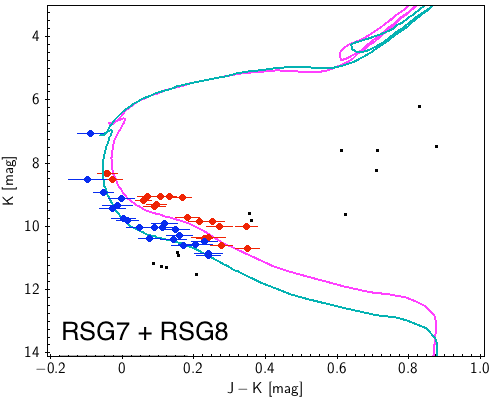}
\end{minipage}\hfill%
\begin{minipage}[t]{0.250\textwidth}\vspace{0pt}
\includegraphics[width=\textwidth]{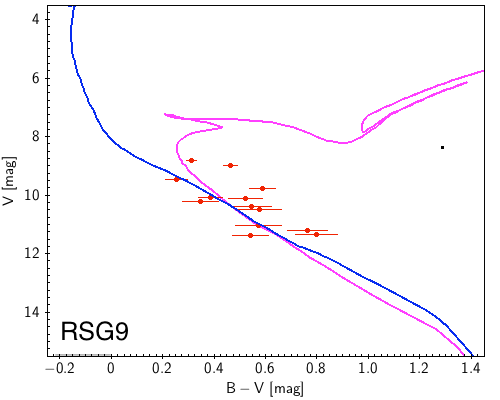}
\end{minipage}\hfill%
\begin{minipage}[t]{0.250\textwidth}\vspace{0pt}
\includegraphics[width=\textwidth]{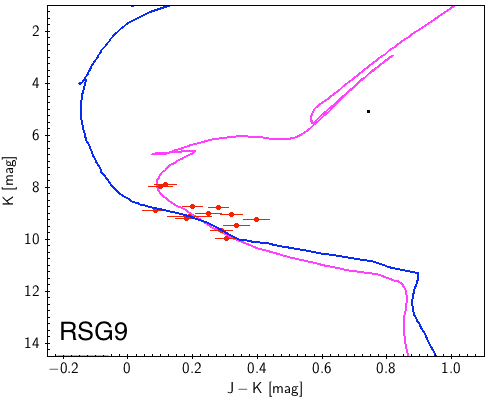}
\end{minipage}\\[5pt]%
\caption{The $(B-V,V)$ and $(J-K_S,K_S)$ colour-magnitude diagrams of the clusters RSG1 to RSG9 from Table~\ref{table:1}.
   Thick red (blue in the case of RSG7) dots with error bars are confirmed photometric members, small black dots
are field stars within the same proper motion slice within a 2-degrees radius (a 3-degrees radius in the case of RSG2) around the cluster centre. Fitted Padova isochrones corresponding
  to the ages given in Table 1 are shown
as coloured lines }\label{Fig1}
\end{figure*}

b) Only 11 Level-2 centres made it into the final process. From the stars in
each Level-2 centre's
{\em sky circles} we determined  a mean position
$\overline{\alpha},\overline{\delta}$ and mean proper motions 
($\overline{\mu_{\alpha}\cos{\delta}}$, $\overline{\mu_{\delta}} $). 
Then we selected an area of 2 degree radius around $\overline{\alpha},\overline{\delta}$ containing the stars
from a proper motion slice around the mean proper motion.

The photometric $B, V$ data from ASCC-2.5 are pretty accurate for stars brighter than $V = 10$, but the accuracy decreases
rapidly for fainter stars. Therefore, we cross-matched the stars around each
Level-2 centre with APASS9 \citep{APASS9} 
to obtain more precise optical photometry
for fainter stars, knowing that there may be small systematic differences between the photometric systems of 
ASCC-2.5 and APASS9. 

In order to determine membership, distance modulus, age and reddening, we
used Padova CMD 2.8\footnote{http://stev.oapd.inaf.it/cgi-bin/cmd} solar metallicity isochrones both within the optical 
($B-V, V$) and the near-infrared ($J-K_S, K_S$) CMDs.
Allowing for possible binarity we consider a star as a probable member if
its offset from the isochrone is less than 2.5-sigma of its photometric error
in magnitude and color in both CMDs. 
\begin{figure*}[t!]
\begin{minipage}[t]{0.24\textwidth}\vspace{0pt}
\includegraphics[width=\textwidth]{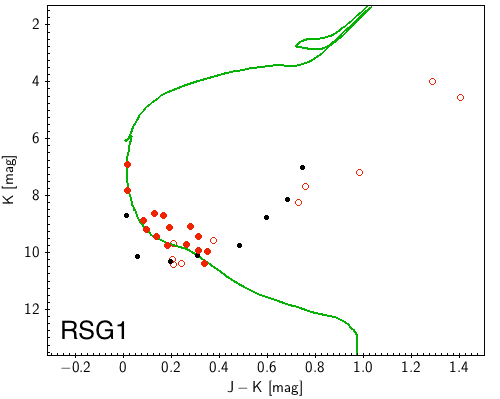}
\end{minipage}\hfill%
\begin{minipage}[t]{0.24\textwidth}\vspace{0pt}
\includegraphics[width=\textwidth]{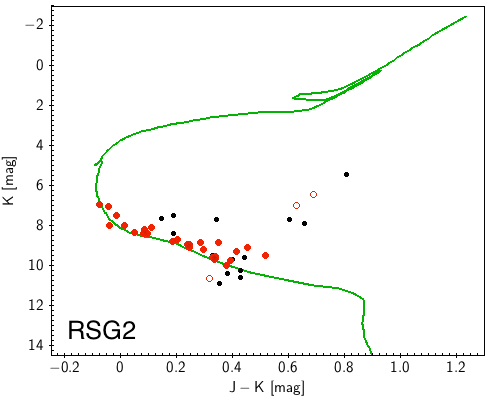}
\end{minipage}\hfill%
\begin{minipage}[t]{0.24\textwidth}\vspace{0pt}
\includegraphics[width=\textwidth]{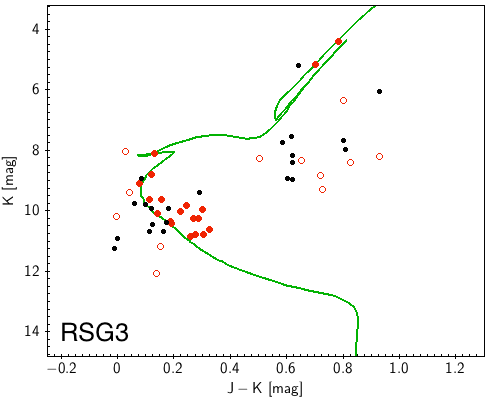}
\end{minipage}\hfill%
\begin{minipage}[t]{0.24\textwidth}\vspace{0pt}
\includegraphics[width=\textwidth]{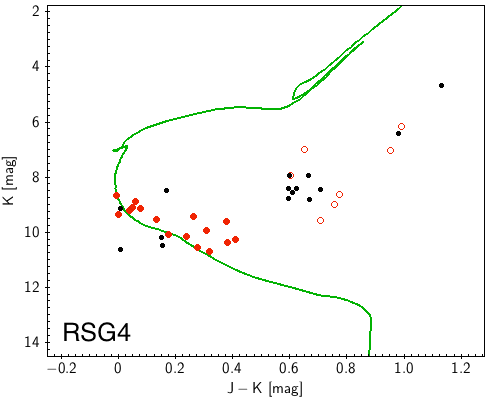}
\end{minipage}\\[5pt]%
\begin{minipage}[t]{0.24\textwidth}\vspace{0pt}
\includegraphics[width=\textwidth]{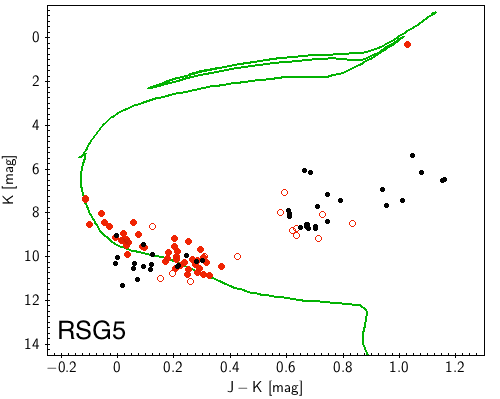}
\end{minipage}\hfill%
\begin{minipage}[t]{0.24\textwidth}\vspace{0pt}
\includegraphics[width=\textwidth]{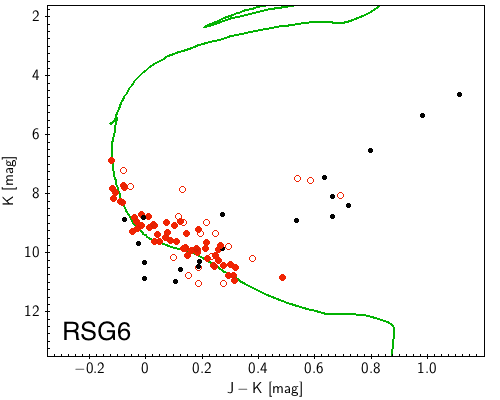}
\end{minipage}\hfill%
\begin{minipage}[t]{0.24\textwidth}\vspace{0pt}
\includegraphics[width=\textwidth]{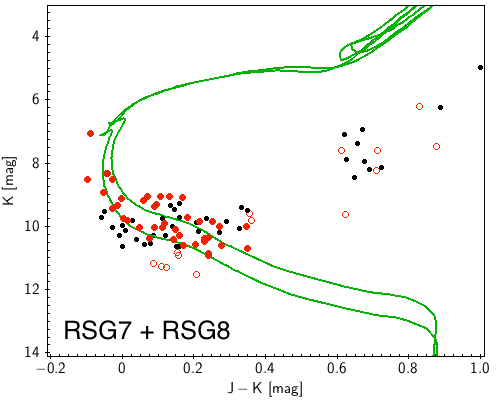}
\end{minipage}\hfill%
\begin{minipage}[t]{0.24\textwidth}\vspace{0pt}
\includegraphics[width=\textwidth]{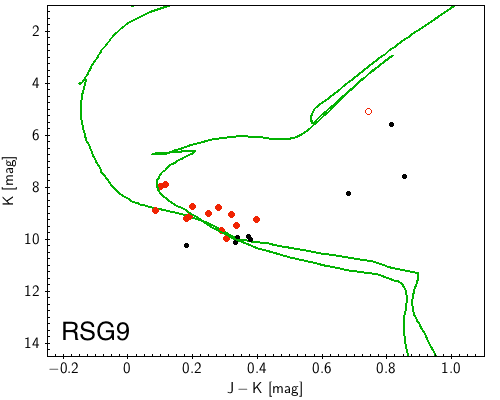}
\end{minipage}\\[5pt]%
\caption{Observed and predicted (by the Besan\c{c}on model)
$(J-K_S,K_S)$ CMDs in the areas of the clusters from Table~\ref{table:1}. Red symbols refer to
observations, where thick dots mark probable members, circles field stars. 
The black dots are predicted field stars from the Besan\c{c}on model of the Galaxy.}
\label{Fig2}
\end{figure*}
\section{Newly found open clusters.} 
For 8 out of 11 Level-2 centres, representing 9 clusters (see below), we obtained reasonable
fits to Padova isochrones.
The newly detected clusterings are located in
the Solar Neighbourhood at distances below 500~pc from the Sun.
The candidates RSG1 to RSG8 are very probably genuine physical groups.
Membership and astrophysical parameters could be determined sufficiently well. 
Nevertheless, accurate parallaxes of at least several reliable cluster stars could
improve the quality of parameter determination.
For RSG9 a definite age cannot be derived, it critically depends on the
secure membership status of the two brightest stars. 

Table~\ref{table:1} summarises the astrophysical parameters of the newly found objects. 
Columns 1 and 2 contain the names of the clusters 
and the number of probable members (N), contained in TYCURAT, as determined in Section 3.2. All the following cluster parameters are determined from these probable members.

The coordinates of
the centres in galactic longitude ($l$),  galactic latitude ($b$), 
 right ascension (RAc) and declination (DEc) are given in  columns 3 to 6. In columns 7 and 8 one
finds the mean proper motion and the precision of the mean motions. The distances D, ages ($\log t$),
the distance Z from the galactic plane, reddening $E(B-V)$ and distance modulus (Dmod) are given
in columns $9 - 13$. The CMDs of the new clusters are shown in Fig.~\ref{Fig1}.
To evaluate a possible contamination by the stellar population
from the general Galactic field  we
used the Besan\c{c}on galactic model \citep{2003A&A...409..523R} with realistic 
extinction included in the simulation, based upon the 3-D extinction maps of \citet{2014ApJ...789...15S}.
As our underlying survey, the Tycho-2 catalogue,
is about 90\% complete down to $V = 11.5$
\citep{tycho2} and rapidly gets incomplete beyond, we cut our simulations at this magnitude.
For each cluster we simulated the contribution of field stars
in the same proper motion range and on an area of $\approx 12.5$ square degrees around the centre 
and show the results in the CMDs of Fig.~\ref{Fig2}.
The number of the predicted stars was always fairly less than the number of the actual stars in the cluster areas, 
a hint to the reality of the over-densities found. 

Each of the nine clusters is briefly descussed below:\\

RSG1\\
The combined use of optical and NIR photometric data allowed
a reliable selection of the most probable members. A parallax 
of $3.8 \pm 0.7$ mas \citep{2007A&A...474..653V} and a spectral type B9
\citep{2014yCat....1.2023S} of the brightest member (HD 32270) are consistent with 
the distance and age determined for RSG1. The members form a
relatively compact group of $1 \times 1$ sq.deg. (5.6 pc $\times$ 5.6 pc)
of young stars moving with a significant tangential velocity of 
19~km/s relative to the Sun.
As seen 
from Fig.~\ref{Fig2} the predicted field-stars contamination is negligible
for this cluster.

RSG2\\
RSG2 is the nearest cluster with the largest proper motion. It groups
around the two B8/B9 stars 19 Lyn A and 19 Lyn B, and includes
three probable  members with a relative parallax accuracy from 11\% to
15\% in \citet{2007A&A...474..653V}. They indicate a mean distance of 197 pc, coinciding well with
our determination based on isochrone fitting. Since RSG2 is so close to us, 
we allowed a search for possible members on a larger 
area of the sky (R < 3~deg).  RSG2 is about as old as the Pleiades but considerably less populated. It may be curious, but it is 
located at about the same galactic longitude ($l=166^{\circ}$ for RSG2 and $l=163^{\circ}$ for 
the Pleiades), but north of the Galactic plane ($b=26^{\circ}$  for RSG2 and $b=-23^{\circ}$ for
the Pleiades). Due to its large proper motion this cluster
is practically free from field star contamination (2 stars obtained from the model). Only when we used a large area of 
200 sq. deg. (factor seven in area) we could show in Fig.~\ref{Fig2} where posible contaminants would lie in the CMD.

RSG3\\
This cluster is rather far away to have accurate trigonometric parallaxes. 
According to \citet{2007A&A...474..653V}, the two brightest members, HD  70358 (K1 III) and HD  70407 (K2 III), 
are located at distances between 270 and 480 pc from the Sun which does not contradict our estimate
from isochrone fitting. RSG3 is the oldest clustering in our sample and it represents
a rather loose group of co-moving stars distributed within an area of 
20~pc~$\times$~20~pc. In the case of RSG3, the stellar population from the Besan\c{c}on model  
illuminates the problem of revealing old clusters in a shallow sky survey. The model gives 
a  contamination  of about 30\% near the
isochrone of the cluster. More accurate astrometric data from the first Gaia data release
will decide upon the fate of this cluster.

RSG4\\
With an age of about 350 Myr, this cluster is a typical representative of the cluster
population  in the Solar Neighbourhood \citep[see][]{2006A&A...445..545P}. A parallax of
$3.5 \pm 0.7$ mas of the brightest member HD 180007 (A0/A0III) is consistent with our estimate of
the  cluster distance. 
According to the model the contamination is rather small,
so we have no doubts on its existence.

RSG5\\
This cluster is the youngest one in our sample. There are four probable members with
parallaxes measured by Hipparcos with a relative accuracy of 10-20\%. The
resulting average distance of about 325 pc ($275 - 435$ pc) is consistent with our 
estimate. The brightest star (31 Cyg) is a spectroscopic binary K4Ib+B4V \citep{2004A&A...424..727P}
with an age of $39.8 \pm 10.3$ Myr \citep{2011MNRAS.410..190T}, which, also, supports our age estimate 
for the cluster. In accordance with the prediction from the model the contamination by field stars is only about 15\%.

\begin{table*}[t!]
\centering
\begin{footnotesize}
\caption{New open clusters.} 
\label{table:1}
\begin{tabular}{|l|c|r|r|r|r|r|r|r|c|r|c|c|}
\hline
\hline
  \multicolumn{1}{|c|}{Name} &
  \multicolumn{1}{c|}{N} &
  \multicolumn{1}{c|}{$l$} &
  \multicolumn{1}{c|}{$b$} &
  \multicolumn{1}{c|}{RAc} &
  \multicolumn{1}{c|}{DEc} &
  \multicolumn{1}{c|}{$\overline{\mu_{\alpha}\cos{\delta}}$ (rms)} &
  \multicolumn{1}{c|}{$\overline{\mu_{\delta}} $ (rms)} &
  \multicolumn{1}{c|}{D} &
  \multicolumn{1}{c|}{$\log t$} &
  \multicolumn{1}{c|}{Z} &
  \multicolumn{1}{c|}{$E(B-V)$} &
  \multicolumn{1}{c|}{Dmod} \\
      &  & deg & deg  & deg & deg & mas/y & mas/y & pc &   & pc & mag &  mag\\
\hline
  RSG1 & 15 & 167.8 & -2.6 & 75.57 & 37.56 & 0.20(0.13) & -12.32(0.13) & 324 & 8.10 & -15 & 0.24 & 7.55\\
  RSG2 & 23 & 162.4 & 26.2  & 110.57 & 54.81 & -2.41(0.13) & -29.54(0.18) & 200 & 8.10 & 88 & 0.04 & 6.50\\
  RSG3 & 19 & 232.0 & 16.1 & 126.12 & -8.81 & -7.98(0.15) & 4.33(0.15) & 437 & 9.00 & 121 & 0.01 & 8.20\\
  RSG4 & 16 & 87.8 &  19.5 & 288.54 & 56.96 & -0.27(0.16) & 5.71(0.17) & 363 & 8.55 & 121 & 0.07 & 7.80\\
  RSG5 & 45 & 81.8 &  6.0 & 303.67 & 45.59 & 3.41(0.09) & 2.22(0.09) & 355 & 7.70 & 37 & 0.04 & 7.75\\
  RSG6 & 54 & 78.9 & 0.6  & 307.49 & 40.08 & 5.86(0.11) & 1.97(0.10) & 347 & 7.80 & 3 & 0.04 & 7.70\\
  RSG7 & 21 & 108.4 & -0.8  & 343.94 & 58.73 & 5.74(0.13) & -0.42(0.16) & 457 & 8.30 & -7 & 0.06 & 8.30\\
  RSG8 & 19 & 108.8 & -0.7  & 344.56 & 59.06 & 5.81(0.19) & -0.70(0.13) & 331 & 8.50 & -4 & 0.05 & 7.60\\
\hline
  RSG9y & 13 & 155.1 & 5.0 & 72.74 & 52.23 & 8.39(0.18) & -15.23(0.19) & 224 & 7.50 & 19 & 0.05 & 6.75\\
  RSG9o & 13 & 155.1 & 5.0 & 72.74 & 52.23 & 8.39(0.18) & -15.23(0.19) & 224 & 9.00 & 19 & 0.02 & 6.75\\
\hline\end{tabular}
\end{footnotesize}
\end{table*}
   \begin{figure}[h!]
   \centering
   \includegraphics[width=0.45\textwidth]{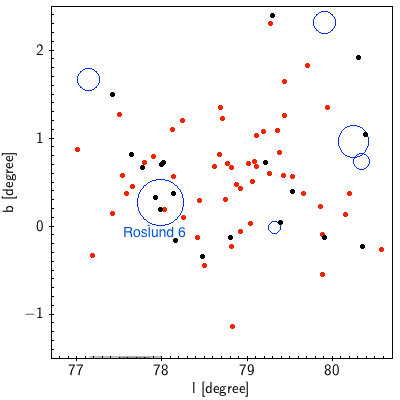}
      \caption{The area around the new cluster RSG6 (galactic coordinates).
      Red dots are the probable members in RSG6, black dots the TYCURAT stars within the same proper motion slice.
      Open clusters from MWSC are shown as large blue circles, the radius corresponding to the cluster radius r2 in MWSC.}
         \label{RSG6area}
   \end{figure}

RSG6\\
As a relatively young, nearby, and well populated cluster, RSG6 is concentrated within an area 
of about 2.5 x 2.5 sq.deg, (15~pc~$\times$~15~pc) covering partly the cluster Roslund~6 \citep{1960PASP...72..205R}, 
which is more distant, 544 pc, older,
$\log t = 8.67$, and smaller, radius 0.3 degree \citep[data from][]{2013A&A...558A..53K}. There are Hipparcos parallaxes
of a relative accuracy better than 18\% for five  B-stars which were selected as probable
RSG6 members. They place the cluster at a distance of about 330 pc ($230 - 450$ pc) which
coincides well with our distance estimate. The spectral type of the brightest star HD 194789, 
B6 IV/V \citep{2014yCat....1.2023S} indicates a young age for RSG6. As RSG6 and Roslund 6 
have nearly the same proper motions (given the precision in \mbox{TYCURAT})
the question if Roslund 6 is either 
a separate cluster behind RSG6 or a part of a large complex RSG6 can only be solved 
definitely with Gaia parallaxes and proper motions. In Fig.~\ref{RSG6area} we show the area 
surrounding Roslund 6 and RSG6 on the sky.
According to the model the contamination is less than 10\%,
so we have no doubts on its existence.

RSG7 and RSG8\\    
The CMDs of the co-moving stars in this sky area reveal a well defined sequence, though it is
too wide to be caused by binarity of several members. A possible explanation is that there 
is a mixture of
several clusterings with about the same proper motions, but at different distances (one behind the other) projected  
on the sky in the same direction. We fit the observed distribution by
two different isochrones, the results are given in Table~\ref{table:1} and in Fig~\ref{Fig1}. Unfortunately, only for the 
brightest star the parallax was measured by Hipparcos with a relative accuracy better than 
20\% resulting in a distance between 355 and 500~pc.  
The Besan\c{c}on model for the area of RSG7 and RSG8 in Fig.~\ref{Fig2} shows moderate
contamination near the sequences of the two clusters. With Gaia data a clear decision can be made
about the number of groups and the membership of stars to one or the other, respectively.

RSG9\\
This group is represented by only 13 co-moving stars distributed over an area of 5 pc $\times$ 10 pc
(with 8 stars concentrated to an inner core of 2 pc $\times$ 4 pc), and with a relatively large
proper motion. 
The major axis of the cluster core is parallel to the galactic plane and the direction
of the proper motion vector coincides with the major axis. 
For one
star a parallax of $3.63 \pm 0.94$ mas \citep{2007A&A...474..653V} was measured by Hipparcos.  Though the 
relative accuracy is rather moderate, this agrees with our distance estimate for RSG9.
The age, however, cannot be determined definitely, i.e. all ages 
between $\log t = 7.5$ and 9.0 are still acceptable. Here, Gaia is needed
for a conclusive membership determination.
On the other hand the contamination by field stars is not an issue for this cluster.
   
   \begin{figure}[h!]
   \centering
   \includegraphics[width=0.45\textwidth]{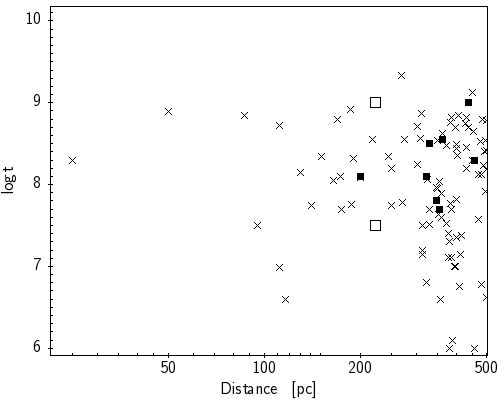}
      \caption{Distances and ages of the newly found clusters within 500 pc.(The new clusters
    are shown as black squares, whereas the MWSC clusters \citep{2013A&A...558A..53K} are shown as
    crosses. The open squares mark the age spread of the cluster RSG9. }
         \label{Figdmodlogt}
   \end{figure}

\section{Discussion}

Our results show that with highly precise proper motions and carefully selected search parameters
it is still possible to reveal hitherto unknown open clusters even in the closer neighbourhood
of the Sun up to 500~pc. On the other hand, we detected no new open cluster beyond that distance.
A possible reason is the low limiting magnitude of Tycho-2. So, the larger the distance modulus and extinction,
the less probable it becomes to get enough un-evolved stars to find a cluster main sequence in the CMDs.
It looks as if this happens at distance modulus 8.5 (roughly 500~pc). 

Due to the shallow depth of Tycho-2 we could find only the brightest and most massive stars in our clusters.
Typical stellar mass functions in open clusters \citep[see, e.g. Figure 3 in][]{2010ARA&A..48..339B} imply
that the total mass of a cluster is
governed by stars around 0.5~$\rm{M_\sun}$  which, at distances prevailing here, are below the limiting magnitude of Tycho-2.
Not knowing the mass we cannot make a statement if the new clusters are still gravitationally bound or
dissolving.

Our findings contribute substantially to the open cluster population within 500~pc
from the Sun. In Fig.~\ref{Figdmodlogt} we show the distribution of the distances and ages of our nine newly found
clusters. The clusters are shown as black squares, only cluster RSG9 is presented as two open squares indicating the range of possible ages.
Also shown are the 92 clusters (crosses) from \citet{2013A&A...558A..53K}.
Although we could only search on 67\% of the sky, our results increase the total number of clusters within 500~pc by 10\%.
Only two of the MWSC clusters, Ruprecht~147 and NGC~752, have $\log t \geq 9$. RSG3 adds one more
cluster to this old cluster generation; RSG9 may add another one if its old age is confirmed.

The first Gaia data release will contain TGAS \citep{2015A&A...574A.115M} supplying proper motions similar to or better than
\mbox{TYCURAT} on the whole sky. But the biggest progress will be achieved by the about 2 million
absolute parallaxes. These will surpass the Hippacos parallaxes in quality and by a factor of
20 in quantity. It will open the path to a substantial leap in open cluster research.

\begin{acknowledgements}
This study was supported by Sonderforschungsbereich SFB 881 "The Milky Way
System" (subprojects B5 and B7) of the German Research Foundation (DFG).
It is a great pleasure to acknowledge Mark Taylor from the Astrophysics Group of the School of Physics at the University of Bristol for
his wonderful work on TOPCAT, Tool for OPerations on Catalogues And Tables. We made extensive use of the major issues of this beautiful tool.
This research has made use of the SIMBAD database and of the VizieR catalogue access tool,
operated at CDS, Strasbourg, France. 
\end{acknowledgements}

%
%
\bibliographystyle{aa}
\bibliography{mybib}

\begin{thebibliography}{26}
\expandafter\ifx\csname natexlab\endcsname\relax\def\natexlab#1{#1}\fi

\bibitem[{{Bastian} {et~al.}(2010){Bastian}, {Covey}, \&
  {Meyer}}]{2010ARA&A..48..339B}
{Bastian}, N., {Covey}, K.~R., \& {Meyer}, M.~R. 2010, \araa, 48, 339

\bibitem[{Dias {et~al.}(2002)Dias, Alessi, Moitinho, \& L{\'e}pine}]{daml02}
Dias, W.~S., Alessi, B.~S., Moitinho, A., \& L{\'e}pine, J. R.~D. 2002, A\&A,
  389, 871

\bibitem[{{Henden} {et~al.}(2016){Henden}, {Templeton}, {Terrell}, {Smith},
  {Levine}, \& {Welch}}]{APASS9}
{Henden}, A.~A., {Templeton}, M., {Terrell}, D., {et~al.} 2016, VizieR Online
  Data Catalog, 2336

\bibitem[{{H{\o}g} {et~al.}(2000){H{\o}g}, {Fabricius}, {Makarov}, {Urban},
  {Corbin}, {Wycoff}, {Bastian}, {Schwekendiek}, \& {Wicenec}}]{tycho2}
{H{\o}g}, E., {Fabricius}, C., {Makarov}, V.~V., {et~al.} 2000, \aap, 355, L27

\bibitem[{{Kharchenko}(2001)}]{2001KFNT...17..409K}
{Kharchenko}, N.~V. 2001, Kinematika i Fizika Nebesnykh Tel, 17, 409

\bibitem[{{Kharchenko} {et~al.}(2005){Kharchenko}, {Piskunov}, {R{\"o}ser},
  {Schilbach}, \& {Scholz}}]{newc109}
{Kharchenko}, N.~V., {Piskunov}, A.~E., {R{\"o}ser}, S., {Schilbach}, E., \&
  {Scholz}, R.-D. 2005, \aap, 440, 403

\bibitem[{{Kharchenko} {et~al.}(2012){Kharchenko}, {Piskunov}, {Schilbach},
  {R{\"o}ser}, \& {Scholz}}]{khea12}
{Kharchenko}, N.~V., {Piskunov}, A.~E., {Schilbach}, E., {R{\"o}ser}, S., \&
  {Scholz}, R.-D. 2012, \aap, 543, A156

\bibitem[{{Kharchenko} {et~al.}(2013){Kharchenko}, {Piskunov}, {Schilbach},
  {R{\"o}ser}, \& {Scholz}}]{2013A&A...558A..53K}
{Kharchenko}, N.~V., {Piskunov}, A.~E., {Schilbach}, E., {R{\"o}ser}, S., \&
  {Scholz}, R.-D. 2013, \aap, 558, A53

\bibitem[{{Mamajek}(2016)}]{2016IAUS..314...21M}
{Mamajek}, E.~E. 2016, in IAU Symposium, Vol. 314, IAU Symposium, ed. J.~H.
  {Kastner}, B.~{Stelzer}, \& S.~A. {Metchev}, 21--26

\bibitem[{{Michalik} {et~al.}(2015){Michalik}, {Lindegren}, \&
  {Hobbs}}]{2015A&A...574A.115M}
{Michalik}, D., {Lindegren}, L., \& {Hobbs}, D. 2015, \aap, 574, A115

\bibitem[{{Piskunov} {et~al.}(2006){Piskunov}, {Kharchenko}, {R{\"o}ser},
  {Schilbach}, \& {Scholz}}]{2006A&A...445..545P}
{Piskunov}, A.~E., {Kharchenko}, N.~V., {R{\"o}ser}, S., {Schilbach}, E., \&
  {Scholz}, R.-D. 2006, \aap, 445, 545

\bibitem[{{Pourbaix} {et~al.}(2004){Pourbaix}, {Tokovinin}, {Batten}, {Fekel},
  {Hartkopf}, {Levato}, {Morrell}, {Torres}, \& {Udry}}]{2004A&A...424..727P}
{Pourbaix}, D., {Tokovinin}, A.~A., {Batten}, A.~H., {et~al.} 2004, \aap, 424,
  727

\bibitem[{{Robin} {et~al.}(2003){Robin}, {Reyl{\'e}}, {Derri{\`e}re}, \&
  {Picaud}}]{2003A&A...409..523R}
{Robin}, A.~C., {Reyl{\'e}}, C., {Derri{\`e}re}, S., \& {Picaud}, S. 2003,
  \aap, 409, 523

\bibitem[{{R{\"o}ser} {et~al.}(2010){R{\"o}ser}, {Demleitner}, \&
  {Schilbach}}]{ppmxl}
{R{\"o}ser}, S., {Demleitner}, M., \& {Schilbach}, E. 2010, \aj, 139, 2440

\bibitem[{{R{\"o}ser} {et~al.}(2011){R{\"o}ser}, {Schilbach}, {Piskunov},
  {Kharchenko}, \& {Scholz}}]{2011A&A...531A..92R}
{R{\"o}ser}, S., {Schilbach}, E., {Piskunov}, A.~E., {Kharchenko}, N.~V., \&
  {Scholz}, R.-D. 2011, \aap, 531, A92

\bibitem[{{Roslund}(1960)}]{1960PASP...72..205R}
{Roslund}, C. 1960, \pasp, 72, 205

\bibitem[{{Schlafly} {et~al.}(2014){Schlafly}, {Green}, {Finkbeiner},
  {Juri{\'c}}, {Rix}, {Martin}, {Burgett}, {Chambers}, {Draper}, {Hodapp},
  {Kaiser}, {Kudritzki}, {Magnier}, {Metcalfe}, {Morgan}, {Price}, {Stubbs},
  {Tonry}, {Wainscoat}, \& {Waters}}]{2014ApJ...789...15S}
{Schlafly}, E.~F., {Green}, G., {Finkbeiner}, D.~P., {et~al.} 2014, \apj, 789,
  15

\bibitem[{{Schmeja} {et~al.}(2014){Schmeja}, {Kharchenko}, {Piskunov},
  {R{\"o}ser}, {Schilbach}, {Froebrich}, \& {Scholz}}]{2014A&A...568A..51S}
{Schmeja}, S., {Kharchenko}, N.~V., {Piskunov}, A.~E., {et~al.} 2014, \aap,
  568, A51

\bibitem[{{Scholz} {et~al.}(2015){Scholz}, {Kharchenko}, {Piskunov},
  {R{\"o}ser}, \& {Schilbach}}]{2015A&A...581A..39S}
{Scholz}, R.-D., {Kharchenko}, N.~V., {Piskunov}, A.~E., {R{\"o}ser}, S., \&
  {Schilbach}, E. 2015, \aap, 581, A39

\bibitem[{{Skiff}(2014)}]{2014yCat....1.2023S}
{Skiff}, B.~A. 2014, VizieR Online Data Catalog, 1

\bibitem[{{Skrutskie} {et~al.}(2006){Skrutskie}, {Cutri}, {Stiening},
  {Weinberg}, {Schneider}, {Carpenter}, {Beichman}, {Capps}, {Chester},
  {Elias}, {Huchra}, {Liebert}, {Lonsdale}, {Monet}, {Price}, {Seitzer},
  {Jarrett}, {Kirkpatrick}, {Gizis}, {Howard}, {Evans}, {Fowler}, {Fullmer},
  {Hurt}, {Light}, {Kopan}, {Marsh}, {McCallon}, {Tam}, {Van Dyk}, \&
  {Wheelock}}]{cat2MASS}
{Skrutskie}, M.~F., {Cutri}, R.~M., {Stiening}, R., {et~al.} 2006, \aj, 131,
  1163

\bibitem[{{Taylor}(2005)}]{2005ASPC..347...29T}
{Taylor}, M.~B. 2005, in Astronomical Society of the Pacific Conference Series,
  Vol. 347, Astronomical Data Analysis Software and Systems XIV, ed.
  P.~{Shopbell}, M.~{Britton}, \& R.~{Ebert}, 29

\bibitem[{{Tetzlaff} {et~al.}(2011){Tetzlaff}, {Neuh{\"a}user}, \&
  {Hohle}}]{2011MNRAS.410..190T}
{Tetzlaff}, N., {Neuh{\"a}user}, R., \& {Hohle}, M.~M. 2011, \mnras, 410, 190

\bibitem[{{van Leeuwen}(2007)}]{2007A&A...474..653V}
{van Leeuwen}, F. 2007, \aap, 474, 653

\bibitem[{{Zacharias} {et~al.}(2015){Zacharias}, {Finch}, {Subasavage},
  {Bredthauer}, {Crockett}, {Divittorio}, {Ferguson}, {Harris}, {Harris},
  {Henden}, {Kilian}, {Munn}, {Rafferty}, {Rhodes}, {Schultheiss}, {Tilleman},
  \& {Wieder}}]{2015AJ....150..101Z}
{Zacharias}, N., {Finch}, C., {Subasavage}, J., {et~al.} 2015, \aj, 150, 101

\bibitem[{{Zacharias} {et~al.}(2013){Zacharias}, {Finch}, {Girard}, {Henden},
  {Bartlett}, {Monet}, \& {Zacharias}}]{2013AJ....145...44Z}
{Zacharias}, N., {Finch}, C.~T., {Girard}, T.~M., {et~al.} 2013, \aj, 145, 44

\end{thebibliography}
\end{document}